\documentclass{article}
\usepackage{spconf,amsmath,graphicx}
\usepackage{multirow, adjustbox}
\usepackage{amsmath,bm}
\usepackage{amssymb}
\usepackage{algorithm}
\usepackage{algpseudocode}
\usepackage{pifont}
\usepackage{xcolor}
\title{Decoupled Spatial and Temporal Processing for Resource Efficient Multichannel Speech Enhancement}
\name{Ashutosh Pandey and Buye Xu}
\address{Meta Reality Labs Research}
\begin{document}
\ninept
\maketitle
\begin{abstract}
We present a novel model designed for resource-efficient multichannel speech enhancement in the time domain, with a focus on low latency, lightweight, and low computational requirements. The proposed model incorporates explicit spatial and temporal processing within deep neural network (DNN) layers. Inspired by frequency-dependent multichannel filtering, our spatial filtering process applies multiple trainable filters to each hidden unit across the spatial dimension, resulting in a multichannel output. The temporal processing is applied over a single-channel output stream from the spatial processing using a Long Short-Term Memory (LSTM) network. The output from the temporal processing stage is then further integrated into the spatial dimension through elementwise multiplication. This explicit separation of spatial and temporal processing results in a resource-efficient network design. Empirical findings from our experiments show that our proposed model significantly outperforms robust baseline models while demanding far fewer parameters and computations, while achieving an ultra-low algorithmic latency of just 2 milliseconds. 
\end{abstract}
\begin{keywords}
Multichannel, lightweight, time-domain, low-latency, low-compute
\end{keywords}
\section{Introduction}
\label{sec:intro}
Multichannel speech enhancement aims at improving the quality and intelligibility of spoken audio in challenging acoustic environments. Applications span from improving voice communication in noisy environments to enhancing automatic speech recognition systems, all relying on robust speech enhancement techniques. The rise of deep learning has led to unprecedented advances in multichannel speech enhancement \cite{wang2017supervised}.

Multichannel speech enhancement through deep neural networks (DNNs) has undergone extensive investigation in recent years, with several prominent approaches emerging. One prevalent method involves the integration of a DNN with a traditional spatial filter, such as the mask-based MVDR beamformer. In this context, the DNN estimates speech and noise statistics to inform the spatial filter's operation \cite{erdogan2016improved, heymann2016neural, gannot2017consolidated, patel2023uxnet, wang2023fsblstm}. Another approach focuses on training DNNs using input features that explicitly encode spatial information \cite{wang2018multi, wang2018combining}. The current mainstream  is centered around either complex spectrum mapping \cite{fu2017complex} or waveform mapping \cite{fu2017raw}. Complex spectral mapping operates in short-time Fourier transform  (STFT) domain and aims at estimating the real and imaginary coefficients of target speech spectrum from the real and imaginary coefficients of noisy speech spectrum \cite{wang2018multi, tolooshams2020channel, tan2022neural, pandey2022multichannel, liu2022drc, lee2023deft}. Waveform mapping, on the other hand, operates in time-domain and directly estimated clean speech samples from noisy speech samples \cite{liu2020multichannel, luo2020end, zhang2020end, pandey2022tparn, pandey2022adhoc}. Despite the substantial performance gains achieved through end-to-end training, traditional spatial filters continue to find wide utility, even in conjunction with strong state-of-the-art models, such as TF-GridNet \cite{wang2023tf, wang2023tfgrid}.

Despite the significant advancements deep learning has ushered into the field, numerous existing models tend to be characterized by substantial computational demands, latency restrictions, and a surplus of parameters. Put differently, DNN models are not particularly resource-efficient. With the increasing prevalence of applications demanding real-time or edge computing, the demand for speech enhancement solutions capable of delivering exceptional performance without overburdening underlying resources has become increasingly urgent. 

Researchers have explored end-to-end training approaches for resource-efficient multichannel speech enhancement. Wang et al. \cite{wang2022stft} introduced a dual-window technique to create a complex spectral mapping model with an algorithmic latency of $4$ ms, further reducing it to $2$ ms through future frame prediction. This concept was recently extended to a strong model utilizing full-band and sub-band recurrent processing \cite{wang2023fsblstm}.

Utilizing waveform mapping to design low-latency models is relatively straightforward, as it takes inspiration from the seminal work of the convolutional time-domain audio separation network (TasNet) \cite{luo2019conv}. This approach involves employing a small frame size and frame shift in the context of end-to-end learning. Specifically, a multichannel convolutional TasNet, employing causal convolutions, was introduced for speech enhancement, achieving an algorithmic latency of $2$ ms. A similar approach was adopted by Patel et. al. \cite{patel2023uxnet}, who proposed a convolutional recurrent model for lightweight, low-compute, and low-latency multichannel speech enhancement. Meanwhile, Pandey et al. \cite{pandey2023llrnn} explored the dual-window concept in time-domain to develop a simple RNN-based model for resource-efficient speech enhancement.

Despite these existing resource-efficient speech enhancement methods, consensus on an acceptable level of computation remains elusive, largely contingent on hardware constraints. It's worth noting that while models in \cite{wang2022stft, wang2023fsblstm, zhang2020end, tu2021two, patel2023uxnet, pandey2023llrnn} excel in ultra-small algorithmic latency, models in \cite{patel2023uxnet, pandey2023llrnn} stand out for their significantly reduced compute requirements.

In this paper, we present a novel approach aimed at enhancing the resource efficiency of speech enhancement models through the decoupling of spatial and temporal processing within DNN layers. Specifically, we introduce a spatio-temporal processing block that leverages spatial processing, akin to frequency-dependent multichannel filtering, to transform a multichannel input signal into another multichannel output signal. Subsequently, temporal processing is performed using a long short-term memory (LSTM) network on one of the channels within the multichannel output. This temporal processing is then efficiently propagated to the remaining channels through elementwise multiplication. The final model is designed by stacking multiple spatio-temporal blocks using dense connections. This straightforward decoupling of spatial and temporal processing not only leads to a more resource-efficient design with reduced computational demands and parameters but also outperforms existing approaches to resource efficient multichannel speech enhancement. 

\section{Model Description}
\label{sec:method}
\subsection{Problem Formulation}
 A multichannel recording $\bm{Y} \in \mathbb{R}^{C \times N}$ captured using an array with $C$ microphones is defined as follows:
\begin{equation}
\bm{Y} = \bm{S}_{d} + \bm{S}_{R} + \bm{N}
\end{equation}
where $N$ represents the number of samples, $\bm{S}_{d}$, $\bm{S}_{r}$, and $\bm{N} \in \mathbb{R}^{C \times N}$, and respectively correspond to direct-path speech, its associated reverberation, and interfering signals recorded by array. The primary goal of multichannel speech enhancement is to produce a reliable estimation, $\bm{s}_{d}^{r}$, of the direct-path speech at a specific reference microphone labeled as $r$, based on the observed noisy recording $\bm{Y}$. In essence, the goal is to effectively eliminate room reverberation and unwanted noises from the degraded speech signal captured at the reference microphone. 

\subsection{Spatial Convolution}
\label{sec:spatial_convolution}
\begin{figure}[!b]
  \centering
  \centerline{\includegraphics[width=\columnwidth]{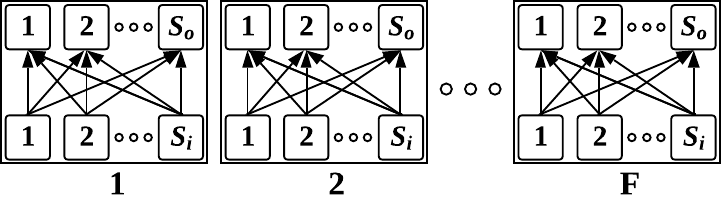}}
\caption{A spatial convolution layer with $S_{i}$ input channels, $S_{o}$ output channels and $F$ hidden units.}
\label{fig:spatial_convolution}
\end{figure}

The proposed model extensively utilizes spatial and temporal processing. The spatial processing is performed using a novel convolution technique known as spatial convolution. Spatial convolution is applied over a given 2D tensor of size $S_{i} \times F$, where $S_{i}$ is viewed as the input spatial dimension and $F$ is viewed as the frequency dimension.  It's important to note that $F$ here does not correspond to the number of frequency bins in the short-time Fourier transform; rather, it refers to the $F$ hidden units within the model, conceptualized as the frequency dimension. In this process, $F$ distinct matrices corresponding to $F$ hidden units, each sized $S_{o} \times S_{i}$, are multiplied to vectors of length $S_{i}$ at corresponding hidden units, resulting in an output tensor of size $S_{o} \times F$ with an output spatial dimension of $S_{o}$. An illustrative diagram of spatial convolution is shown in Fig. \ref{fig:spatial_convolution}. This approach draws inspiration from frequency-dependent multichannel filtering, where distinct filters are employed at each frequency bin. Furthermore, by configuring the number of output channels to a value exceeding one, the model gains the ability to learn multiple spatial filters. It's noteworthy that implementing this layer is straightforward, achievable through either grouped convolutions or \emph{einsum} within existing deep learning libraries.
\label{ssec:seq_bf}

\begin{figure}[htb]
  \centering
  \centerline{\includegraphics[width=\columnwidth]{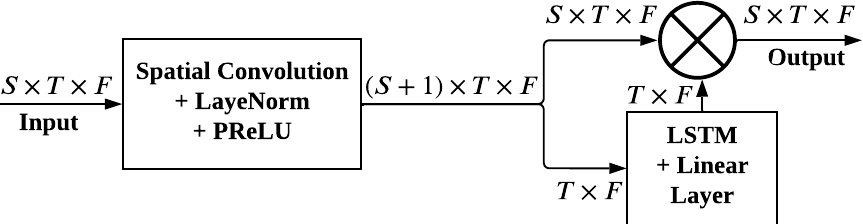}}
\caption{Spatio-temporal Block.}
\label{fig:spatio_temporal_block}
\end{figure}
\subsection{Spatio-Temporal Block}
We propose a novel building block, illustrated in Fig. \ref{fig:spatio_temporal_block}, comprising spatial convolution for spatial processing and LSTMs for temporal processing. The processing begins with an input tensor of dimensions $S \times T \times F$, where it is conceptualized as having $T$ frames, each with spatial width $S$ and frequency width $F$.

First, we apply a spatial convolution layer with $S+1$ output channels, followed by layer normalization \cite{ba2016layer} and parametric rectified linear unit(PReLU) nonlinearity \cite{he2015delving}. This operation yields an output tensor of dimensions ($S+1) \times T \times F$. Within this output, the first channel undergoes processing through an LSTM layer with a hidden size of $F$, followed by a linear layer with size $F$. Notably, the LSTM operates across time, facilitating efficient temporal processing over a single channel, rather than processing all channels concurrently.

In the final step, the output of the temporal processing stage is elementwise multiplied with the remaining $S$ channels from the output of spatial processing. This intricate process serves to propagate temporal refinement over the spatial dimension. 

This unique decoupling of spatial and temporal processing results in a parameter and compute-efficient processing block tailored for multichannel speech enhancement.

\begin{figure*}[!t]
  \centering
  \centerline{\includegraphics[width=\textwidth]{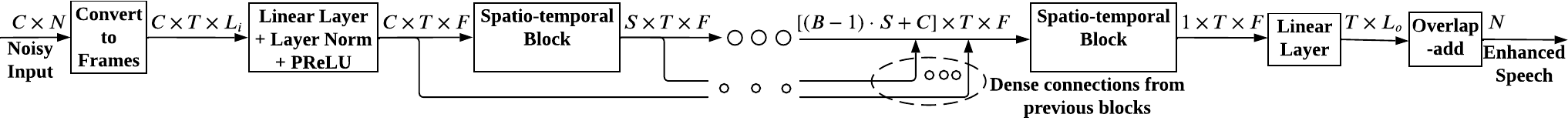}}
\caption{The proposed model for time-domain multichannel speech enhancement.}
\label{fig:dllrnn}
\end{figure*}

\subsection{Model Architecture}
The architectural design of the proposed model is visualized in Fig. \ref{fig:dllrnn}. Initially, an input signal of dimensions $C \times N$ is transformed into overlapping frames sized $C \times T \times L_i$, where $T$ signifies the number of frames, and $L_i$ denotes the input frame size. These frames of size $L_i$ are subsequently converted into a latent representation of size $F$. This conversion process involves employing a linear layer followed by layer normalization and PReLU activation. In this context, $F$ characterizes the frequency dimension of the network as discussed in Section \ref{sec:spatial_convolution}.

The resulting output undergoes a series of transformations through a stack of $B$ spatio-temporal blocks, each yielding $S$ output channels. In this context, $S$ corresponds to the spatial dimension of the network. Each spatio-temporal block leverages outputs from all preceding blocks, forming connections akin to densely connected convolutional neural networks. These dense connections are strategically incorporated to enhance feature learning by promoting more effective gradient signal propagation throughout the network. Furthermore, to enhance model parameters and computational efficiency, dense connections are integrated by concatenating signals along the spatial (channel) dimension.

The final spatio-temporal block uses one output channel to convert a multichannel tensor to a single-channel tensor. This output is then projected to size $L_{o}$ using a linear layer to generate enhanced frames. It is noteworthy that $L_{i}$ can potentially exceed $L_{o}$, and in such cases, the network outputs the rightmost $L_{o}$ samples within the input frame, ensuring an algorithmic latency of $L_{o}$. To achieve this, the input is padded with $L_{i} - L_{o}$ zeros at the beginning \cite{pandey2023llrnn}.  

\subsection{Loss Function}
All the models are trained using the phase-constrained magnitude (PCM) loss, which was initially introduced in \cite{pandey2021dense} and is outlined in Eq. \ref{eq:pcm} as follows:
\begin{equation}
\label{eq:pcm}
    L_{PCM}(\bm{x}, \hat{\bm{x}}) = L_{SM}(\bm{x}, \hat{\bm{x}}) +  L_{SM}(\bm{y - x}, \bm{y} - \hat{\bm{x}})
\end{equation}

Here, the $L_{SM}$ term is defined as:

\begin{equation}
\begin{aligned}
L_{SM}(\bm{x}, \hat{\bm{x}}) = \frac{1}{T \cdot F}\sum_{t=1}^{T}\sum_{f=1}^{F}|&(|X(t, f){r}| + |X(t, f){i}|) - \\
& (|\hat{X}(t, f){r}| + |\hat{X}(t, f){i}|)|
\end{aligned}
\end{equation}

In this equation, $X_{r}(t, f)$ and $X_{i}(t, f)$ respectively represent the real and imaginary components of the spectral coefficient at frequency bin $f$ of the $t_{th}$ frame.The variable $T$ represents the number of frames, while $F$ denotes the number of frequency bins. The $L_{PCM}$ loss employs $L_{1}$ distance between the $L_1$ norms of estimated and target spectral coefficients for both speech and interference components.

\section{Experiments}
\subsection{Dataset}
We utilize the Interspeech2020 DNS Challenge corpus (Reddy et al., 2020) as our data source for generating pairs of clean and noisy signals. In the training set, all speakers are randomly divided into groups comprising training, test, and validation speakers, with a distribution ratio of 85\%, 5\%, and 10\%, respectively. Similarly, we categorize noise sources into distinct sets for training, test, and validation purposes.

To enable the creation of multichannel data, we deploy an eight-microphone circular array with a radius of $10$ cm. Our data generation process, similar to the methodologies employed in \cite{pandey2022tparn, pandey2022multichannel, pandey2022adhoc, pandey2023llrnn}, is described in Algorithm \ref{algo:spatialization}. We generate 80K training samples, 1.6K validation samples, and 3.2K test samples, each consisting of 10-second-long utterances.

We utilize the Pyroomacoustics library in Python, employing the image method with an order of $6$. The absorption coefficient for reverberation is uniformly sampled from the range $[0.1, 0.4]$. The signal-to-noise ratio (SNR) is computed by summing the energy of the direct path across all channels for the signal energy and summing the interference energy (excluding speech reverberation) across all channels for the noise energy.
\begin{algorithm}[!b]
\caption{Multichannel data generation algorithm.}
\label{algo:spatialization}
\begin{algorithmic}
\footnotesize
\For {\emph{split} in \{train, test, validation \}}
\For {speech utterances in \emph{split} }
\begin{itemize}
\item Draw room length and width from [$3$,$10$] m, and height from [$2$, $5$] m
\item Draw $1$ array location and $1$ speech source location;
\item Get $8$ uniformly placed mic locations on a circle of radius $10$ cm centered at array location
\item Draw $N_{ns}$ number of noise sources uniformly from [$1$, $10$]
\item Draw $N_{ns}$ random noise locations inside room
\item Generate RIRs corresponding to the speech source location and $N_{ns}$ noise locations for mic locations in circular array
\item Draw $N_{ns}$ noise utterances from noises in \emph{split} 
\item Propagate speech and noise signals to mics by convolving with corresponding RIRs
\item Draw a value $snr$ from [-10, 10] dB, and add speech and noises at each mic using a scale so that the SNR is $snr$
\end{itemize}
\EndFor
\EndFor
\end{algorithmic}
\end{algorithm} 
\begin{table}
    \centering
    \begin{adjustbox}{width=\columnwidth}
    \begin{tabular}{|c|c|c|c|c|c|}
\cline{2-6}
\multicolumn{1}{c|}{}& \ \ STOI \ \  & \ \ PESQ \ \   & SI-SDR &  GFLOPs & Params.(M)\\
\cline{1-6}
Unprocessed & 65.8 & 1.63 & -7.5 & - & - \\
\hline
D-LL-RNN-64-1-8 & 82.6 & 2.34 & 3.7 & 0.90 & 0.34 \\
D-LL-RNN-64-2-8 & 83.5 & 2.40 & 3.7 & 0.93 & 0.34 \\
D-LL-RNN-64-4-8 & 84.0 & 2.40 & 4.0 & 1.01 & 0.38 \\
D-LL-RNN-64-8-8 & 84.6 & 2.45 & 4.1 & 1.25 & 0.49 \\
D-LL-RNN-64-8-6 & 83.7 & 2.39 & 3.7 & 0.95 & 0.34 \\
D-LL-RNN-64-8-4 & 81.9 & 2.29 & 3.0 & 0.69 & 0.22 \\
D-LL-RNN-32-8-8 & 81.0 & 2.24 & 2.1 & \textbf{0.48} & \textbf{0.17} \\
D-LL-RNN-128-8-8 & 87.4 & 2.60 & 5.8 & 3.67 & 1.57 \\
D-LL-RNN-200-4-8 & 89.0 & 2.75 & 6.8 & 7.06 & 3.14 \\
D-LL-RNN-256-4-8 & 89.5 & 2.79 & 7.2 & 11.10 & 5.05 \\
D-LL-RNN-256-8-8 & \textbf{89.9} & \textbf{2.83} & \textbf{7.5} & 12.06 & 5.50 \\
\hline
LL-RNN-128-2ms & 80.8 & 2.27 & 2.9 & 1.34 & 0.44 \\
LL-RNN-200-2ms & 83.9 & 2.43 & 4.2 & 2.78 & 1.03 \\
LL-RNN-256-2ms & 85.6 & 2.51 & 4.9 & 4.25 & 1.66 \\
LL-RNN-300-2ms & 86.2 & 2.56 & 5.3 & 5.61 & 2.26 \\
LL-RNN-400-2ms & 87.5 & 2.64 & 6.0 & 9.40 & 3.97 \\
LL-RNN-512-2ms & 88.3 & 2.69 & 6.5 & 14.79 & 6.46 \\
MC-Conv-Tasnet-2ms& 86.3 & 2.57 & 5.6 & 10.32 & 5.13 \\
MC-CRN-2ms & 84.0 & 2.38 & 3.9 & 6.73 & 2.32 \\
MC-CRN-4ms & 85.7 & 2.51 & 4.7 & 6.73 & 2.32 \\
UXNet-128-2ms & 77.3 & 2.10 & 1.1 & 0.67 & 0.21 \\
UXNet-256-2ms & 80.9 & 2.25 & 2.9 & 2.12 & 0.81 \\
FSB-LSTM-4ms & 88.2 & 2.68 & 5.8 & 7.80 & 1.97 \\
\hline
\end{tabular}
\end{adjustbox}
    \caption{Comparisons between D-LL-RNN and baseline models.} 
    \label{tbl:results}
\end{table}

\subsection{Experimental Settings}
All the utterances are resampled to $16$ kHz. A given multichannel waveform input is normalized (multiplied with a scalar) to have an overall variance of one across all the microphones. The hop size for converting signals to frames is set to $16$ samples ($1$ ms). The input and output frame sizes, $L_{i}$ and $L_o$, are respectively set to $256$ samples ($16$ ms) and $32$ samples ($2$ ms), i.e., the model is trained for an algorithmic latency of $2$ ms. We perform ablation experiments to assess effectiveness of hyperparameters $S$, $F$ and $B$. 

All the models are developed, trained and evaluated using PyTorch. All of the layer-normalization modules in the proposed model normalize the last (frequency) dimension. We train models for $200$ epochs using random chunks of $4$ seconds cropped out of $10$ seconds long utterances with a batch size of $16$. The Adam optimizer \cite{kingma2014adam} with $amsgrad=True$ and a constant learning rate of $0.0002$ is used. The gradient norm is clipped to a value of $0.03$. A combination of automatic mixed precision (AMP) and Nvidia V100 GPUs is utilized for a much faster training.

All the models are evaluated using short-time objective intelligibility (STOI) \cite{taal2011algorithm}, perceptual evaluation of speech quality (PESQ) \cite{rix2001perceptual} and scale-invariant signal-to-distortion ratio (SI-SDR). The direct-path speech at the first microphone is used as the reference to compute all the metrics. Average scores over $3.2$K test utterances are reported. The amount of computation is reported in Giga FLOPs for processing one second of 8-channel speech.

\subsection{Baseline Models}
We evaluate our proposed system against several low-latency multichannel speech enhancement methods. Initially, we train three distinct time-domain baseline models: the low-latency RNN (LL-RNN) introduced in \cite{pandey2023llrnn}, the multichannel convolutional time-domain audio separation network (MC-Conv-TasNet) from references \cite{zhang2020end, tu2021two}, and the UX-Net model presented in \cite{patel2023uxnet}. All these time-domain models are trained to function with an algorithmic latency of 2 milliseconds. The LL-RNN model is trained using different widths, denoted as $H$ in the original paper. This approach allows us to evaluate and compare the model's performance under various computational conditions. Following the approach outlined in the original paper \cite{patel2023uxnet}, the UX-Net is trained with two different values for the hyperparameter $N$.  

In addition to the time-domain models mentioned earlier, we also develop and train two frequency-domain models. The first one is the multichannel convolutional recurrent network (MC-CRN) proposed in \cite{wang2022stft}. The second is the full-band sub-band LSTM (FSBLSTM) network introduced in \cite{wang2023fsblstm}. These models perform complex spectral mapping, enhancing both the real and imaginary parts of the complex spectrum. To achieve low latency, they employ a dual-window approach, resulting in a latency of $4$ ms. We also train the MC-CRN model using future frame prediction, reducing its latency to $2$ ms, following the approach outlined in \cite{wang2022stft}.

\subsection{Experimental Results}
In Table \ref{tbl:results}, we provide objective scores for all the models. Our proposed model is labeled as D-LL-RNN, where "D" signifies "decoupled," and "LL-RNN" stands for low-latency RNN. Additionally, we include hyperparameter values denoted by the format D-LL-RNN-$F$-$S$-$B$. For clarity, we present all baseline models along with their respective algorithmic latencies as suffixes. 

Firstly, it is noticeable that with fixed values of $F = 64$ and $B = 8$, the model's performance shows gradual improvement as we increment the value of S. Specifically, by increasing $S$ from $1$ to $8$, we achieve a $2$\% improvement in STOI, a $0.11$ increase in PESQ, and a $0.4$ dB improvement in SI-SDR. In this scenario, the computational workload increases by 0.35 GFLOPs, and the number of parameters in the model increases by $0.15$ million (M). 

Furthermore, we notice a decline in performance when we decrease the value of $B$. However, the most significant impact on both computational resources and performance is observed when altering the value of $F$. For instance, when we double $F$ from $64$ to $128$, we see a substantial improvement in STOI by $2.8$\%, PESQ by $0.15$, and SI-SDR by $1.7$ dB. This performance boost comes at a cost, as computational requirements nearly double, and the number of parameters in the model increases by $2.2$ times.

In our final comparison with baseline models, we can observe some interesting findings. For instance, the model labeled D-LLRNN-$64$-$8$-$4$, which utilizes only $0.69$ GFLOPs and $0.22$ M parameters, outperforms LLRNN-$128$, which consumes $1.34$ GFLOPs and has $0.44$ M parameters. Notably, UX-Net-$128$, despite using a similar amount of computational resources and parameters, lags significantly behind. It is outperformed by $4.6$\% in STOI, $0.19$ in PESQ, and $1.9$ dB in SI-SDR.

Similarly, D-LL-RNN-$128$-$8$-$8$ surpasses LLRNN-$300$, MC-CRN, and MC-Conv-TasNet while being considerably more efficient in terms of computational resources and the number of parameters. The best-performing baseline model, FSB-LSTM, is highly resource-intensive, utilizing a substantial $7.8$ GFLOPs. What's intriguing is that D-LL-RNN-$200$-$4$-$8$, with a smaller computational footprint of $7.1$ GFLOPs, even outperforms FSB-LSTM. Notably, performance can be further enhanced by increasing values of $F$ and $S$, with the highest scores achieved by D-LL-RNN-$256$-$8$-$8$.

In summary, D-LL-RNN offers a versatile framework that allows for various ways to balance the trade-off between computational resources and performance. It consistently outperforms existing approaches while maintaining efficiency. 
\section{Conclusions}
We have introduced and extensively evaluated a novel approach of decoupled spatial and temporal processing inside a DNN model for multichannel speech enhancement. The proposed models consistently outperforms existing approaches, even with fewer resources and a streamlined architecture, making it a promising solution for real-world multichannel speech enhancement applications. This research advances the integration of low-latency processing and high-quality enhancement, paving the way to more efficient solutions in the field.





\bibliographystyle{IEEEbib}
\bibliography{mybib}

\end{document}